\documentstyle[11pt,paspconf,epsf]{article}

\begin{document}

\title{Optical Galaxy Selection}
\author{Stacy McGaugh}
\affil{Department of Astronomy, University of Maryland, USA}

\begin{abstract}
Our view of the properties of galaxies is strongly affected by the way
in which we survey for them.  I discuss some aspects of selection effects
and methods to compensate for them.  One result is an estimate of the
surface brightness distribution.  I believe this is progress, but
considerable uncertainty remains.
\end{abstract}


\section{Limits and Selection Effects}

A basic goal of galaxy research is to identify and characterize the galaxy
populations which inhabit the universe.  Astronomers undertake large surveys
to discover galaxies, but this is only part of the battle.  In order to
transform from the observed, apparent distributions of galaxy properties
and recover the desired intrinsic distributions, one must
understand and correct for the selection characteristics of galaxy
surveys.

An essential fact about galaxies, as they appear on survey plates, is that
they are not point sources.  Some of the consequences of this simple fact
were first quantified by Disney (1976), who pointed out that the intrinsic
distribution could be very different from the apparent one.  At the least,
it is necessary to generalize the formal procedures developed for point sources
(Disney \& Phillipps 1983; McGaugh et al.\ 1995).

The appearance of galaxies
requires many parameters to approximate: they have multiple components
(bulge and disk); they have spiral arms and dust lanes; their morphology
can be a strong function of wavelength depending on their recent star formation
history; etc.  So, we simplify.  Here I will pretend that a tolerable
approximation can be had with axially averaged radial surface brightness
profiles.  This reduces the dimensionality of the problem
immensely.  The surface brightness profiles of galaxies obtained in this way
are tolerably described as a combination of bulge and disk (4 parameters)
or with a generalized (S\'ersic 1969) profile (3 parameters).  For brevity,
I will reduce this further to 2 parameters by limiting the discussion to
galaxies dominated by exponential disks:
\begin{equation}
\mu(r) = \mu_0 + 1.086 \frac{r}{h}.
\end{equation}
This reduces complex galaxies to a characteristic central surface
brightness $\mu_0$ and scale size $h$.
In some cases this is a fair approximation.

\begin{figure}
\plotone[37 54 522 692]{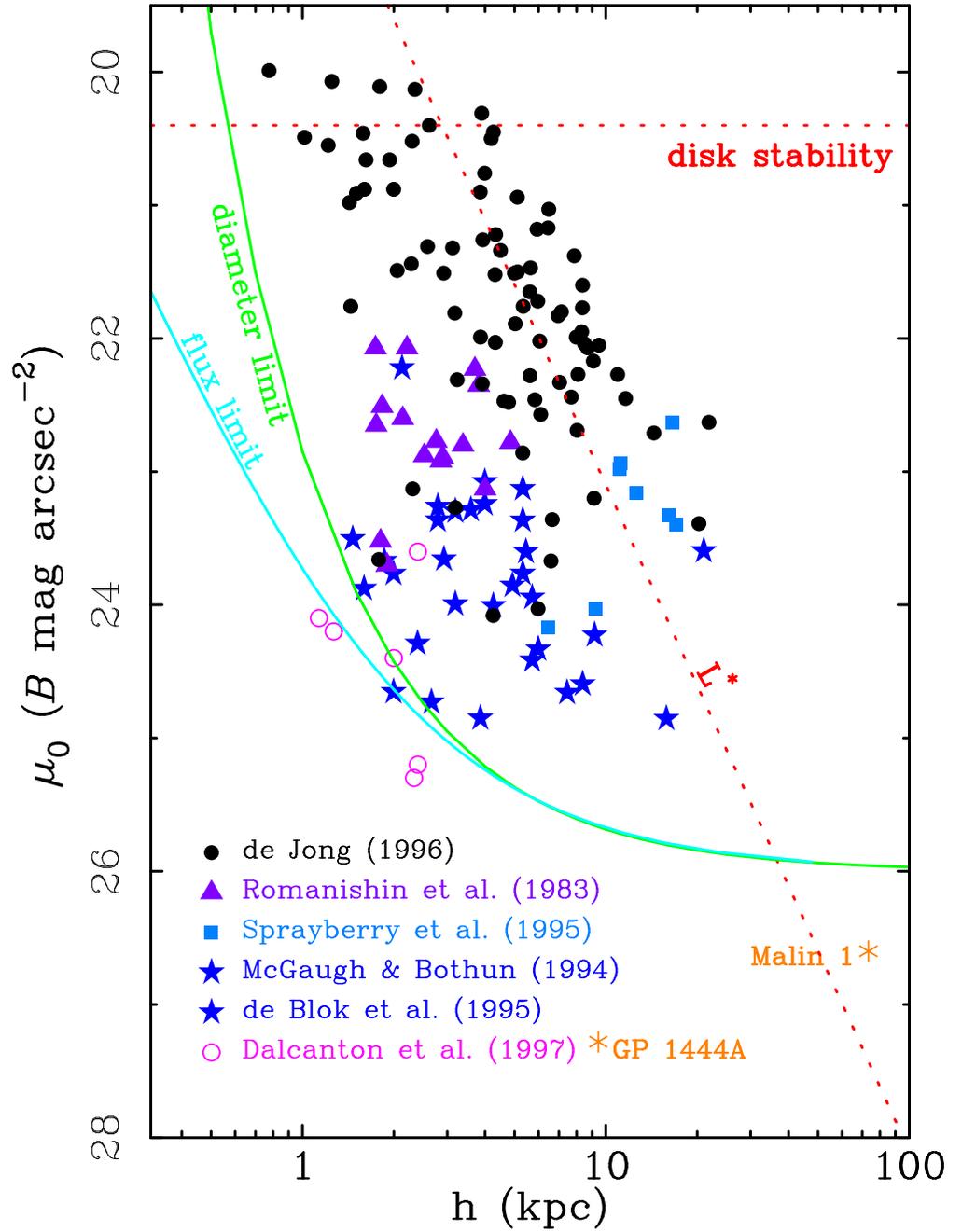}
%\plotone{mcgaughs1.ps}
\caption{The distribution of disk galaxies in the central surface
brightness-scale length plane.  Disks exist at all ($\mu_0,h$) up to
maxima in $\mu_0$ and $L$.  Lower bounds are imposed by observational
selection limits.  Illustrative selection lines are drawn for
$\mu_{\ell} = 26$, typical of good photographic plates.} \label{SHsel}
\end{figure}

Galaxies inhabit a large range in surface brightness and scale length
(Figure 1).  Real limits appear at high surface brightness and luminosity.
Here, objects are prominent and are easily selected by surveys.  The
limit at $L^*$ represents the well known turndown in galaxy numbers at
this point.  The limit at high $\mu_0$ may represent a physical
restriction based on the requirement that disks be stable (Milgrom 1989;
McGaugh \& de Blok 1998), though a massive bulge or dark halo
could in principle stabilize
even higher surface brightness disks.  At the dim and small end, there is
no clearly defined physical limit: our knowledge here is circumscribed by
selection effects.  Galaxies which are intrinsically small or low surface
brightness are hard to find, regardless of how common they may be.
Examples of very low surface brightness galaxies are known to exist,
implying that the sparsely populated space in Figure 1 is just waiting
to be filled.

In addition to a survey diameter or flux (magnitude) limit, there must (at the
least) be a second survey parameter quantifying the limiting isophote
$\mu_{\ell}$ at which these are measured (Disney 1976).
These are shown in Figure 1 by extracting from equation 1 the
radius or flux of a galaxy at the limiting isophote:
\begin{equation}
r > r_{\ell} = 0.92 h (\mu_{\ell}-\mu_0)
\end{equation}
for diameter selection and
\begin{equation}
m < m_{\ell} = \mu_0 - 5\log h -2.5\log[f(\mu_0,\mu_{\ell})] + C
\end{equation}
for selection by apparent magnitude.  See McGaugh et al.\ (1995) for the
definition of $f(x)$.

The problem becomes obvious:
to reach small scale lengths is difficult; to find galaxies dimmer than the
isophotal limit is impossible.  However, the situation is even worse than
illustrated in Figure 1, where it is assumed that galaxies can be detected
right down to the formal survey limit.  Even maintaining this fiction, one
is really battling the volume sampling function $V(\mu_0,h)$ (Disney \&
Phillipps 1983; McGaugh et al.\ 1995) illustrated in Figure 2.  This is
much more restrictive than the lines in Figure 1 imply.

\section{Volume Sampling and Intrinsic Properties}

The volume sampled by any given survey is a strong function of both the
survey parameters ($\mu_{\ell},r_{\ell}$ or $m_{\ell}$) and the intrinsic
properties of galaxies ($\mu_0,h$).  A progressively larger volume is sampled
for intrinsically larger and higher surface brightness galaxies.  Such
galaxies will always dominate the apparent numbers of objects in catalogs
regardless of the intrinsic bivariate distribution $\Phi(\mu_0,h)$
(McGaugh 1996).

\begin{figure}[t]
\plottwo{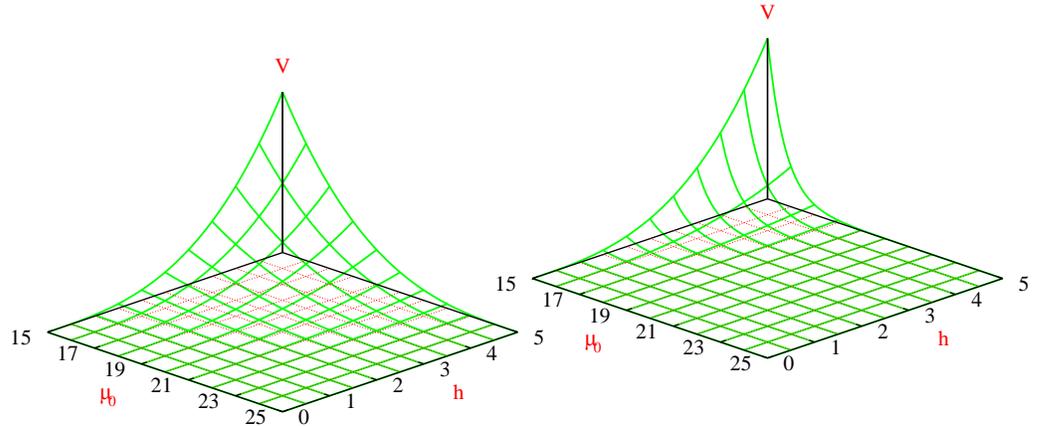}{MBS.flux.cps}
%\plottwo{mcgaughs2a.ps}{mcgaughs2b.ps}
\vspace{-1in}
\caption{The volume sampled by (a) diameter and (b) flux limited surveys as
a function of surface brightness and scale length.}
\label{diamsel}
\end{figure}

It is possible, in principle, to correct for $V(\mu_0,h)$ and recover
the intrinsic distribution from the observed one.  A good example is
given by de Jong (1996 and these proceedings).  Here I restrict myself to
the projection of the bivariate distribution $\Phi(\mu_0,h)$
along the surface brightness axis, which I call the surface brightness
distribution $\Phi(\mu_0)$ (Figure 3).  I derive this by {\it assuming\/}
$\mu_0$ and $h$ are uncorrelated (see Figure 1).  While
almost certainly not true in detail, this procedure does not produce a result
which is grossly at odds with the results of the more complete analysis
of de Jong.

\begin{figure}[t]
\plotone[33 404 543 715]{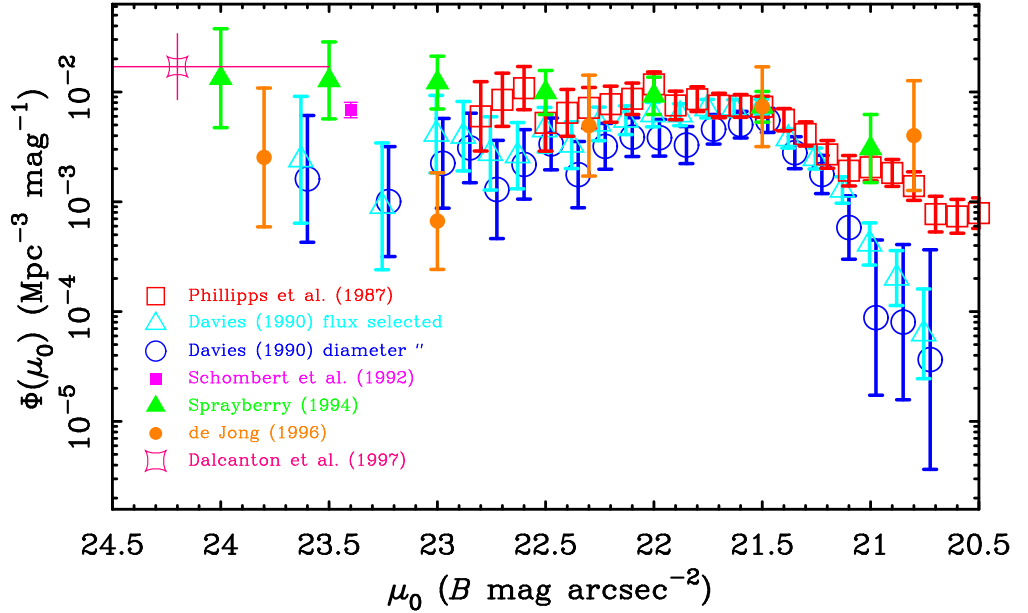}
%\plotone{mcgaughs3.ps}
\caption{The surface brightness distribution, i.e., the number density of
galaxies at each central surface brightness.  There is a sharp feature
at the bright end corresponding to the Freeman value, perhaps corresponding
to a threshold for disk stability.  At the dim end large numbers of
low surface brightness galaxies are indicated.  There is considerable
uncertainty in $\Phi(\mu_0)$ at all surface brightnesses.  Neither the shape
nor the precise value of the Freeman limit are well determined.
} \label{absnumdens}
\end{figure}

The surface brightness distribution is broad, being {\it roughly\/} flat below
the Freeman (1970) value.  Low surface brightness galaxies are numerous, with
approximately equal numbers of galaxies in each bin of central surface
brightness.  The surface brightness distribution falls off rather more
steeply at the bright end.

If the assumption made about the independence of $\mu_0$ and $h$ fails,
it has a predictable effect on the shape of $\Phi(\mu_0)$.  If galaxies
become systematically smaller as surface brightness decreases (as some
unwritten lore presumes), then the surface brightness distribution will
{\it rise\/} towards lower surface brightnesses.  Since both surface brightness
and size act against such dim, small galaxies, the required volume correction
is larger than made here.  On the other hand, if galaxies tend to be larger
at lower surface brightnesses, the surface brightness distribution will fall
more rapidly than shown.
There is some hint of such a trend in Figure 1 (see also McGaugh \&
de Blok 1997).  However, we base all this on incomplete
data which is fraught with the very selection effects for which we would
like to correct!

\begin{figure}[t]
\plotone[33 404 543 715]{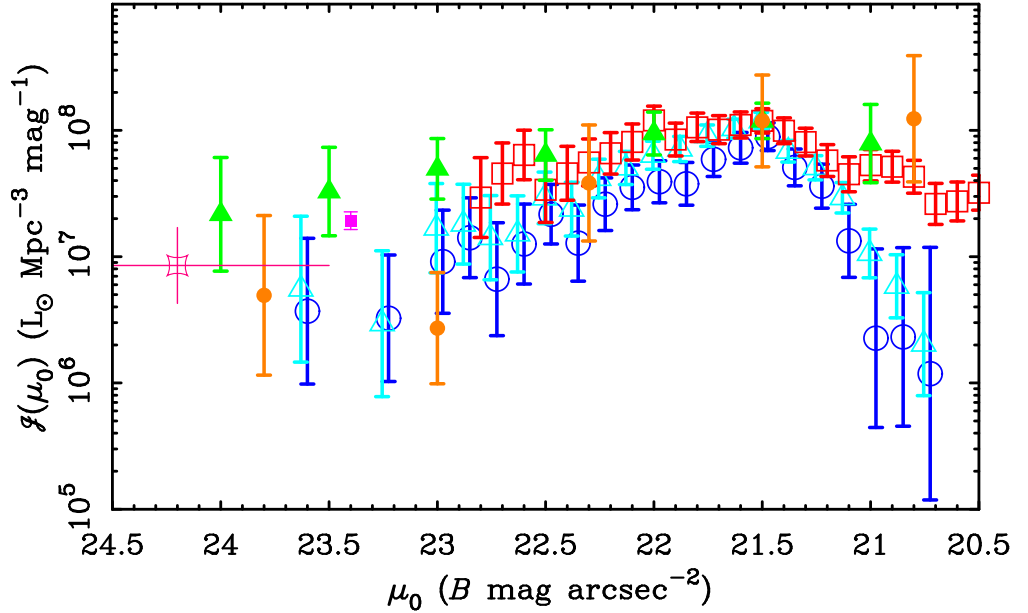}
%\plotone{mcgaughs4.ps}
\caption{Similar to the previous figure, but now the luminosity density.
The luminosity density declines with dropping surface brightness.  This
happens simply because lower surface brightness galaxies are fainter
{\it on average\/} than high surface brightness galaxies.  There is a
significant fraction of the total luminosity density due to all disks in the
low surface brightness tail ($< 30\%$), but it seems unlikely that
a large or even divergent population remains hidden.} \label{abslumdens}
\end{figure}

Different data in Figure 3 are consistent, given the large uncertainties.
I do see some potential for controversy at the dim end.  In an HI selected
sample, Zwaan (these proceedings)
reports a sharp turn down faintwards of $\mu_0 \approx 24$.
Yet Dalcanton et al.\ (1997) not only detect galaxies dimmer
than this, but also claim to measure a rather high density of them.
One possibility is
that the HI column densities of very low surface brightness galaxies
are so low that the HI is ionized by the extragalactic UV radiation field.
However, I do not think this can work.  Only slightly brighter low surface
brightness galaxies have abundant HI (McGaugh \& de Blok 1997; Schombert et al.\
1997), and the column densities are only a little lower than in high surface
brightness spirals (de Blok et al.\ 1996).  Moreover, by this reasoning, the
$10^{11} {\cal M}_{\sun}$ of neutral hydrogen in Malin 1 should all be ionized.

That low surface brightness galaxies are {\it numerous\/} does not
necessarily mean that they contribute a large
amount to the integrated luminosity density (Figure 4).  If the assumption
I have made holds, low surface brightness galaxies are on average the
same size as high surface brightness galaxies, and are therefore
less luminous.  The luminosity density thus declines with surface brightness.

This estimate of the luminosity density (Figure 4) as a function of
surface brightness is more robust than the number density (Figure 3).
If low surface brightness galaxies are larger on average than high surface
brightness galaxies, they give away less in luminosity.  However, they
must then be less numerous in Figure 3, and the effects tend to offset.
They do not offset perfectly (McGaugh 1996); because of the severity of
the volume sampling function, it is easier to conceal luminosity in the small,
dim population than in Malin-1-like giants.

\section{Complications}

While I have emphasized the severity of selection effects and the magnitude
of the volume corrections required, I have so far actually painted a fairly
rosy picture.  I have made simple assumptions about the
properties of $\Phi(\mu_0,h)$ and $V(\mu_0,h)$, and have further assumed
that one can simply apply the volume sampling correction to galaxy samples
which are complete as advertised.  In reality, none of these things are likely
to be realized, and there are a number of other important effects
which have been completely neglected.

To mention a just few:
\begin{itemize}
\item Galaxies are not axially symmetric exponential disks.
\item The galaxy SED matters for both bandpass and redshift effects.
\item There are systematic measurement, as well as selection, effects.
\item Inclination and internal extinction matter.
\item Large scale structure may render meaningless a `fair' volume.
\end{itemize}
While it is convenient to assume the exponential profile as the next best
thing to pretending galaxies are point sources, there is no guarantee that
this is adequate for purposes of selection.
Early type galaxies are generally bulge dominated.  Though
the disk may well dominate in the regions of the selection isophote, the
steep inner profile of the bulge makes these galaxies more prominent than
assumed, especially to flux limited surveys.  For late type galaxies the
exponential profile is a better description of the radial profile, but these
galaxies tend to be irregular and have large deviations from axial symmetry.
These effects tend to exacerbate further the already large difference in
accessible volume between bright and dim galaxies.  Moreover,
features like spiral arms, star forming regions, and dust lanes can
complicate matters in a way which is difficult to generalize.

Another issue is the actual spectral energy distribution (SED) of galaxies.
While I have assumed a generic $\mu_0$ in the survey band which defines
$\mu_{\ell}$, one would really like to know the bolometric surface brightness.
As $\mu_0^{bol} \rightarrow \mu_{\ell}$, the visibility of a galaxy depends
quite sensitively on color
(McGaugh 1996).  This may be a partial explanation for why the
multi-band CCD survey of O'Neil et al.\ (1997) turns up more red low surface
brightness galaxies than had previous surveys conducted on blue sensitive
photographic plates (Schombert et al.\ 1992; Impey et al.\ 1996).  If typical
galaxy redshifts for a survey are high enough for $K$-corrections to be
important, things are even more complicated (Ferguson, these proceedings).

One crucial thing in trying to recover the galaxy luminosity function
is the systematic underestimate of the luminosities of low surface brightness
galaxies made by measuring at some limiting isophote (McGaugh 1994;
Dalcanton 1998).  This is a particularly pernicious effect, because it is
severe only for a very small percentage of galaxies in a complete sample.
Yet these
are precisely the objects for which large volume corrections are necessary.
This magnifies systematic errors which in turn have a
disproportionate effect on the derived luminosity function.  All fit parameters
($L^*$, $\phi^*$, $\alpha$) can be affected, but the faint end slope is
particularly susceptible to underestimation. Worse,
many analyses still assume galaxies are point sources, which is tantamount
to drawing the flux selection boundary in Figure 1 as a straight line
parallel to that for $L^*$.

So far, I have assumed all galaxies are face on.  This is bad, but not
too terrible {\it if\/} galaxies are optically thin.  This is probably the
case for low surface brightness galaxies (McGaugh 1994; Tully et al.\ 1998).
It is probably not the case for high surface brightness
spirals.  Just how optically thick spirals may be remains controversial.
For the purposes of selection, internal extinction can have a significant
inclination dependent effect on the position of a galaxy in Figure 1
(see Lu 1998).

Finally, I'd like to mention that a fundamental assumption
is that we can probe a large enough volume of the universe
for it to be considered homogeneous and representative.  Yet as
redshift surveys press ever deeper, we discover ever larger features, with no
clear convergence as yet (Huchra, these proceedings).  While the selection
effects embodied in the volume sampling function will always be important,
it is not obvious that large scale structure will admit a straightforward
correction.  Yet again, this problem is particularly severe at the dim end.
While large surveys are sensitive to high surface brightness $L^*$ galaxies
over impressive volumes, they still probe only modest volumes for small and
low surface brightness galaxies.

\section{Conclusions}

In spite of some of the daunting problems, substantial progress has been made.
There is great scope for further improvement.  If any of the difficulties
I have mentioned imposes some fundamental limitation, we are as yet far from
reaching it.

\acknowledgments  I am grateful to many of the attendees of this meeting
for discussing these topics over the years.  I would particularly like to
thank Jon Davies for organizing a great conference.

\end{document}